\documentclass[preprint,superscriptaddress,nofootinbib]{revtex4}
  \usepackage{graphicx}
  \begin{document}
  \title{The ${\Upsilon}(nS)$ ${\to}$ $B_{c}D_{s}$, $B_{c}D_{d}$
         decays with perturbative QCD approach}
  \author{Junfeng Sun}
  \affiliation{Institute of Particle and Nuclear Physics,
              Henan Normal University, Xinxiang 453007, China}
  \author{Yueling Yang}
  \affiliation{Institute of Particle and Nuclear Physics,
              Henan Normal University, Xinxiang 453007, China}
  \author{Qingxia Li}
  \affiliation{Institute of Particle and Nuclear Physics,
              Henan Normal University, Xinxiang 453007, China}
  \author{Haiyan Li}
  \affiliation{Institute of Particle and Nuclear Physics,
              Henan Normal University, Xinxiang 453007, China}
  \author{Qin Chang}
  \affiliation{Institute of Particle and Nuclear Physics,
              Henan Normal University, Xinxiang 453007, China}
  \author{Jinshu Huang}
  \affiliation{College of Physics and Electronic Engineering,
              Nanyang Normal University, Nanyang 473061, China}
  \begin{abstract}
  The ${\Upsilon}(nS)$ ${\to}$ $B_{c}D_{s}$, $B_{c}D_{d}$
  weak decays are studied with the pQCD approach firstly.
  It is found that branching ratios
  ${\cal B}r({\Upsilon}(nS){\to}B_{c}D_{s})$
  ${\sim}$ ${\cal O}(10^{-10})$ and
  ${\cal B}r({\Upsilon}(nS){\to}B_{c}D_{d})$
  ${\sim}$ ${\cal O}(10^{-11})$, which might be measurable in
  the future experiments.
  \end{abstract}
  \maketitle

  \section{Introduction}
  \label{sec01}
  Since the discovery of bottomonium (the bound states of the bottom
  quark $b$ and the corresponding antiquark $\bar{b}$, i.e., $b\bar{b}$)
  at Fermilab in 1977 \cite{herb,innes},
  remarkable achievements have been made in the understanding
  of the properties of bottomonium, thanks to the endeavor from
  the experiment groups of CLEO, BaBar, Belle, CDF, D0, LHCb,
  ATLAS, and so on \cite{1212.6552}.
  The upsilon, ${\Upsilon}(nS)$, is the $S$-wave spin-triplet
  state, $n^{3}S_{1}$, of bottomonium with the well established
  quantum number of $I^{G}J^{PC}$ $=$ $0^{-}1^{--}$ \cite{pdg}.
  The typical total widths of the upsilons below the
  kinematical open-bottom threshold (where the radial
  quantum number $n$ $=$ 1, 2 and 3) are a few tens of
  keV (see Table \ref{tab:bb}),
  at least two orders of magnitude lower less than those of
  bottomonium above the $B\bar{B}$ threshold.
  (note that for simplicity, the notation ${\Upsilon}(nS)$ will
  denote the ${\Upsilon}(1S)$, ${\Upsilon}(2S)$ and ${\Upsilon}(3S)$
  mesons in the following content if not specified explicitly)
  As it is well known, the ${\Upsilon}(nS)$ meson decays primarily
  through the annihilation of the $b\bar{b}$ pairs into three
  gluons, which are suppressed by the phenomenological
  Okubo-Zweig-Iizuka rule \cite{o,z,i}.
  The allowed $G$-parity conserving transitions,
  ${\Upsilon}(nS)$ ${\to}$ ${\pi}{\pi}{\Upsilon}(mS)$ and
  ${\Upsilon}(nS)$ ${\to}$ ${\eta}{\Upsilon}(mS)$ where
  $3$ $\ge$ $n$ $>$ $m$ ${\ge}$ $1$,
  are greatly limited by the compact phase spaces,
  because the mass difference $m_{{\Upsilon}(3S)}$ $-$
  $m_{{\Upsilon}(2S)}$ is just slightly larger than $2m_{\pi}$,
  and $m_{{\Upsilon}(2S)}$ $-$ $m_{{\Upsilon}(1S)}$
  is just slightly larger than $m_{\eta}$.
  The coupling strengths of the electromagnetic and radiative
  interactions are proportional to the electric charge of the
  bottom quark, $Q_{b}$ $=$ $-1/3$ in the unit of
  ${\vert}e{\vert}$.
  Besides, the ${\Upsilon}(nS)$ meson can also decay
  via the weak interactions within the standard model,
  although the branching ratio is small,
  about $2/{\tau}_{B}{\Gamma}_{{\Upsilon}}$
  ${\sim}$ ${\cal O}(10^{-8})$ \cite{pdg},
  where ${\tau}_{B}$ and ${\Gamma}_{{\Upsilon}}$
  are the lifetime of the $B_{u,d,s}$ meson and the
  total width of the ${\Upsilon}(nS)$ meson, respectively.
  In this paper, we will study the ${\Upsilon}(nS)$ ${\to}$
  $B_{c}D_{s}$, $B_{c}D_{d}$ weak decays with the
  perturbative QCD (pQCD) approach \cite{pqcd1,pqcd2,pqcd3}.
  The motivation is listed as follows.

   \begin{table}[h]
   \caption{Summary of the mass, total width and data samples
   of the ${\Upsilon}(1S,2S,3S)$ mesons.}
   \label{tab:bb}
   \begin{ruledtabular}
   \begin{tabular}{ccccc}
    & \multicolumn{2}{c}{properties \cite{pdg}}
    & \multicolumn{2}{c}{data samples ($10^{6}$) \cite{1406.6311}} \\ \cline{2-3} \cline{4-5}
      meson & mass (MeV) & width (keV) & Belle & BaBar \\ \hline
   ${\Upsilon}(1S)$ & $9460.30{\pm}0.26$ & $54.02{\pm}1.25$
                    & $102{\pm}2$ & ...   \\
   ${\Upsilon}(2S)$ & $10023.26{\pm}0.31$ & $31.98{\pm}2.63$
                    & $158{\pm}4$ & $98.3{\pm}0.9$ \\
   ${\Upsilon}(3S)$ & $10355.2{\pm}0.5 $ & $20.32{\pm}1.85$
                    & $11{\pm}0.3$ &  $121.3{\pm}1.2$
   \end{tabular}
   \end{ruledtabular}
   \end{table}

  From the experimental point of view,
  (1)
  over $10^{8}$ ${\Upsilon}(nS)$ data samples have been
  accumulated by the Belle detector at the KEKB and
  the BaBar detector at the PEP-II  $e^{+}e^{-}$
  asymmetric energy colliders \cite{1406.6311} (see Table \ref{tab:bb}).
  It is hopefully expected that
  more and more upsilons will be collected with great
  precision at the running upgraded LHC and the forthcoming
  SuperKEKB.
  An abundant data samples offer a realistic possibility to
  search for the ${\Upsilon}(nS)$ weak decays which in some
  cases might be detectable.
  (2)
  The signals for the ${\Upsilon}(nS)$ ${\to}$ $B_{c}D_{s,d}$
  weak decays should be clear and easily
  distinguishable from background, because
  the back-to-back final states with opposite
  electric charges have definite momentums and energies in
  the center-of-mass frame of the ${\Upsilon}(nS)$ meson.
  In addition, the identification of either a single flavored
  $D_{s,d}$ or $B_{c}$ meson can be used not only to avoid
  the low double-tagging efficiency \cite{zpc62.271}, but
  also to provide an unambiguous evidence of
  the ${\Upsilon}(nS)$ weak decay.
  It should be noticed that on one hand, the ${\Upsilon}(nS)$
  weak decays are very challenging to be observed experimentally
  due to their small branching ratios, on the other hand,
  any evidences of an abnormally large production rate of
  either a single charmed or bottomed meson might be a hint
  of new physics beyond the standard model \cite{zpc62.271}.

  From the theoretical point of view,
  the ${\Upsilon}(nS)$ weak decays permit one to cross check
  parameters obtained from the $B$ meson decays,
  to further explore the underlying dynamical mechanism of
  the heavy quark weak decay, to test various theoretical
  approaches and to improve our understanding on the
  factorization properties.
  Phenomenologically, the ${\Upsilon}(nS)$ ${\to}$
  $B_{c}D_{s}$, $B_{c}D_{d}$ weak decays are favored
  by the color factor due to the external $W$ emission
  topological structure, and by the Cabibbo-Kabayashi-Maskawa
  (CKM) elements ${\vert}V_{cb}{\vert}$ due to the $b$
  ${\to}$ $c$ transition, so usually their branching
  ratio should not be too small. In addition,
  these two decay modes are the $U$-spin partners with each
  other, so the flavor symmetry breaking effects can be
  investigated.
  However, as far as we know, there is no study concerning
  on the ${\Upsilon}(nS)$ ${\to}$ $B_{c}D_{s,d}$
  weak decays theoretically and experimentally at the moment.
  We wish this paper can provide a ready reference to the
  future experimental searches.
  Recently, many attractive methods have been fully developed
  to evaluate the hadronic matrix elements (HME) where
  the local quark-level operators are sandwiched between
  the initial and final hadron states,
  such as the pQCD approach \cite{pqcd1,pqcd2,pqcd3},
  the QCD factorization \cite{qcdf1,qcdf2,qcdf3}
  and the soft and collinear effective theory
  \cite{scet1,scet2,scet3,scet4},
  which could give an appropriate explanation for
  many measurements on the nonleptonic $B_{u,d}$ decays.
  In this paper, we will estimate the branching ratios
  for the ${\Upsilon}(nS)$ ${\to}$ $B_{c}D_{s,d}$ weak
  decays with the pQCD approach to offer a possibility
  of searching for these processes at the future experiments.

  This paper is organized as follows.
  Section \ref{sec02} is devoted to the theoretical framework
  and the amplitudes for the ${\Upsilon}(nS)$ ${\to}$
  $B_{c}D_{s,d}$ decays.
  We present the numerical results and discussion in section \ref{sec03},
  and conclude with a summary in the last section.

  \section{theoretical framework}
  \label{sec02}
  \subsection{The effective Hamiltonian}
  \label{sec0201}
  Using the operator product expansion and renormalization
  group equation, the effective Hamiltonian responsible for
  the ${\Upsilon}(nS)$ ${\to}$ $B_{c}D_{s,d}$
  weak decays is written as \cite{9512380}
   \begin{equation}
  {\cal H}_{\rm eff}\, =\,
   \frac{G_{F}}{\sqrt{2}}\,
   \sum\limits_{q=d,s}
   \Big\{ V_{cb} V_{cp}^{\ast}
   \sum\limits_{i=1}^{2}
   C_{i}({\mu})\,Q_{i}^{q}({\mu})
  -V_{tb} V_{tp}^{\ast}
   \sum\limits_{j=3}^{10}
   C_{j}({\mu})\,Q_{j}^{q}({\mu}) \Big\}
   + {\rm H.c.}
   \label{hamilton},
   \end{equation}
  where $G_{F}$ $=$ $1.166{\times}10^{-5}\,{\rm GeV}^{-2}$ \cite{pdg}
  is the Fermi coupling constant;
  the CKM factors are expressed as a power series in
  the Wolfenstein parameter ${\lambda}$ ${\sim}$ $0.2$ \cite{pdg},
  \begin{eqnarray}
  V_{cb}V_{cs}^{\ast} &=&
  +            A{\lambda}^{2}
  - \frac{1}{2}A{\lambda}^{4}
  - \frac{1}{8}A{\lambda}^{6}(1+4A^{2})
  +{\cal O}({\lambda}^{7})
  \label{eq:ckm01}, \\
  V_{tb}V_{ts}^{\ast} &=& -V_{cb}V_{cs}^{\ast}
  - A{\lambda}^{4}({\rho}-i{\eta})
  +{\cal O}({\lambda}^{7})
  \label{eq:ckm02},
  \end{eqnarray}
 for the ${\Upsilon}(nS)$ ${\to}$ $B_{c}D_{s}$ decays, and
  \begin{eqnarray}
  V_{cb}V_{cd}^{\ast} &=& -A{\lambda}^{3}
  +{\cal O}({\lambda}^{7})
  \label{eq:ckm03}, \\
  V_{tb}V_{td}^{\ast} &=& +A{\lambda}^{3}(1-{\rho}+i{\eta})
  + \frac{1}{2}A{\lambda}^{5}({\rho}-i{\eta})
  +{\cal O}({\lambda}^{7})
  \label{eq:ckm04}.
  \end{eqnarray}
  for the ${\Upsilon}(nS)$ ${\to}$ $B_{c}D_{d}$ decays.
  The Wilson coefficients $C_{i}(\mu)$ summarize the
  physical contributions above the scale of ${\mu}$,
  and have been reliably calculated to the next-to-leading
  order with the renormalization group assisted
  perturbation theory.
  The local operators are defined as follows.
    \begin{eqnarray}
    Q_{1}^{q} &=&
  [ \bar{c}_{\alpha}{\gamma}_{\mu}(1-{\gamma}_{5})b_{\alpha} ]
  [ \bar{q}_{\beta} {\gamma}^{\mu}(1-{\gamma}_{5})c_{\beta} ]
    \label{q1}, \\
    Q_{2}^{q} &=&
  [ \bar{c}_{\alpha}{\gamma}_{\mu}(1-{\gamma}_{5})b_{\beta} ]
  [ \bar{q}_{\beta}{\gamma}^{\mu}(1-{\gamma}_{5})c_{\alpha} ]
    \label{q2},
    \end{eqnarray}
    \begin{eqnarray}
    Q_{3}^{q} &=& \sum\limits_{q^{\prime}}
  [ \bar{q}_{\alpha}{\gamma}_{\mu}(1-{\gamma}_{5})b_{\alpha} ]
  [ \bar{q}^{\prime}_{\beta} {\gamma}^{\mu}(1-{\gamma}_{5})q^{\prime}_{\beta} ]
    \label{q3}, \\
    Q_{4}^{q} &=& \sum\limits_{q^{\prime}}
  [ \bar{q}_{\alpha}{\gamma}_{\mu}(1-{\gamma}_{5})b_{\beta} ]
  [ \bar{q}^{\prime}_{\beta}{\gamma}^{\mu}(1-{\gamma}_{5})q^{\prime}_{\alpha} ]
    \label{q4}, \\
    Q_{5}^{q} &=& \sum\limits_{q^{\prime}}
  [ \bar{q}_{\alpha}{\gamma}_{\mu}(1-{\gamma}_{5})b_{\alpha} ]
  [ \bar{q}^{\prime}_{\beta} {\gamma}^{\mu}(1+{\gamma}_{5})q^{\prime}_{\beta} ]
    \label{q5}, \\
    Q_{6}^{q} &=& \sum\limits_{q^{\prime}}
  [ \bar{q}_{\alpha}{\gamma}_{\mu}(1-{\gamma}_{5})b_{\beta} ]
  [ \bar{q}^{\prime}_{\beta}{\gamma}^{\mu}(1+{\gamma}_{5})q^{\prime}_{\alpha} ]
    \label{q6},
    \end{eqnarray}
    \begin{eqnarray}
    Q_{7}^{q} &=& \sum\limits_{q^{\prime}} \frac{3}{2}Q_{q^{\prime}}\,
  [ \bar{q}_{\alpha}{\gamma}_{\mu}(1-{\gamma}_{5})b_{\alpha} ]
  [ \bar{q}^{\prime}_{\beta} {\gamma}^{\mu}(1+{\gamma}_{5})q^{\prime}_{\beta} ]
    \label{q7}, \\
    Q_{8}^{q} &=& \sum\limits_{q^{\prime}} \frac{3}{2}Q_{q^{\prime}}\,
  [ \bar{q}_{\alpha}{\gamma}_{\mu}(1-{\gamma}_{5})b_{\beta} ]
  [ \bar{q}^{\prime}_{\beta}{\gamma}^{\mu}(1+{\gamma}_{5})q^{\prime}_{\alpha} ]
    \label{q8}, \\
    Q_{9}^{q} &=& \sum\limits_{q^{\prime}} \frac{3}{2}Q_{q^{\prime}}\,
  [ \bar{q}_{\alpha}{\gamma}_{\mu}(1-{\gamma}_{5})b_{\alpha} ]
  [ \bar{q}^{\prime}_{\beta} {\gamma}^{\mu}(1-{\gamma}_{5})q^{\prime}_{\beta} ]
    \label{q9}, \\
    Q_{10}^{q} &=& \sum\limits_{q^{\prime}} \frac{3}{2}Q_{q^{\prime}}\,
  [ \bar{q}_{\alpha}{\gamma}_{\mu}(1-{\gamma}_{5})b_{\beta} ]
  [ \bar{q}^{\prime}_{\beta}{\gamma}^{\mu}(1-{\gamma}_{5})q^{\prime}_{\alpha} ]
    \label{q10},
    \end{eqnarray}
  where $Q_{1,2}^{q}$, $Q_{3,{\cdots},6}^{q}$, and $Q_{7,{\cdots},10}^{q}$
  are usually called as the tree operators, QCD
  penguin operators, and electroweak penguin operators,
  respectively;
  ${\alpha}$ and ${\beta}$ are color indices;
  $q^{\prime}$ denotes all the active quarks at the scale of
  ${\mu}$ ${\sim}$ ${\cal O}(m_{b})$, i.e.,
  $q^{\prime}$ $=$ $u$, $d$, $s$, $c$, $b$;
  and $Q_{q^{\prime}}$ is the electric charge
  of the $q^{\prime}$ quark in the unit of ${\vert}e{\vert}$.

  \subsection{Hadronic matrix elements}
  \label{sec0202}
  Theoretically, to obtain the decay amplitudes, the
  remaining essential work and also the most complex part
  is the calculation of the hadronic matrix elements
  of local operators as accurate as possible.
  Combining the $k_{T}$ factorization theorem \cite{npb366}
  with the collinear factorization hypothesis,
  and based on the Lepage-Brodsky approach for exclusive processes \cite{prd22},
  the HME can be written as the convolution of universal
  wave functions reflecting the nonperturbative contributions
  with hard scattering subamplitudes containing the
  perturbative contributions within the pQCD framework,
  where the transverse momentums of quarks are retained
  and the Sudakov factors are introduced, in order to
  regulate the endpoint singularities and provide a
  naturally dynamical cutoff on the nonperturbative
  contributions \cite{pqcd1,pqcd2,pqcd3}.
  Generally, the decay amplitude can be separated into three
  parts: the Wilson coefficients $C_{i}$ incorporating the
  hard contributions above the typical scale of $t$,
  the process-dependent scattering amplitudes $T$ accounting
  for the heavy quark decay, and the universal wave functions
  ${\Phi}$ including the soft and long-distance contributions,
  i.e.,
  \begin{equation}
  {\int} dx\, db\,
  C_{i}(t)\,T(t,x,b)\,{\Phi}(x,b)e^{-S}
  \label{hadronic},
  \end{equation}
  where $x$ is the longitudinal momentum fraction of valence
  quarks, $b$ is the conjugate variable of the transverse
  momentum $k_{T}$, and $e^{-S}$ is the Sudakov factor.

  \subsection{Kinematic variables}
  \label{sec0203}
  In the center-of-mass frame of the ${\Upsilon}(nS)$ mesons,
  the light cone kinematic variables are defined as follows.
  \begin{eqnarray}
  p_{{\Upsilon}} &=& p_{1}\, =\, \frac{m_{1}}{\sqrt{2}}(1,1,0)
  \label{kine-p1}, \\
  p_{B_{c}} &=& p_{2}\, =\, (p_{2}^{+},p_{2}^{-},0)
  \label{kine-p2}, \\
  p_{D_{s,d}} &=& p_{3}\, =\, (p_{3}^{-},p_{3}^{+},0)
  \label{kine-p3}, \\
  k_{i} &=& x_{i}\,p_{i}+(0,0,\vec{k}_{iT})
  \label{kine-ki}, \\
  {\epsilon}_{\Upsilon}^{\parallel} &=& \frac{1}{ \sqrt{2} }(1,-1,0)
  \label{kine-1el}, \\
  p_{i}^{\pm} &=& (E_{i}\,{\pm}\,p)/\sqrt{2}
  \label{kine-pipm}, \\
  s &=& 2\,p_{2}{\cdot}p_{3}
  \label{kine-s}, \\
  t &=& 2\,p_{1}{\cdot}p_{2} = 2\,m_{1}\,E_{2}
  \label{kine-t}, \\
  u &=& 2\,p_{1}{\cdot}p_{3} = 2\,m_{1}\,E_{3}
  \label{kine-u},
  \end{eqnarray}
  \begin{equation}
  p = \frac{\sqrt{ [m_{1}^{2}-(m_{2}+m_{3})^{2}]\,[m_{1}^{2}-(m_{2}-m_{3})^{2}] }}{2\,m_{1}}
  \label{kine-pcm},
  \end{equation}
  where $x_{i}$ is the longitudinal momentum fraction;
  $\vec{k}_{iT}$ is the transverse momentum;
  $p$ is the common momentum of final states;
  ${\epsilon}_{\Upsilon}^{\parallel}$ is the
  longitudinal polarization vector of the
  ${\Upsilon}(nS)$ meson;
  $m_{1}$ $=$ $m_{{\Upsilon}(nS)}$,
  $m_{2}$ $=$ $m_{B_{c}}$ and
  $m_{3}$ $=$ $m_{D_{s,d}}$ denote the masses
  of the ${\Upsilon}(nS)$, $B_{c}$ and $D_{s,d}$
  mesons, respectively.
  The notation of momentum is displayed in Fig.\ref{fig:fey}(a).

  \subsection{Wave functions}
  \label{sec0204}
  With the notation of Refs. \cite{jhep0605,jhep0703},
  the HME of diquark operators squeezed between
  the vacuum and the ${\Upsilon}(nS)$,
  $B_{c}$, $D_{q}$ mesons are defined as follows.
  \begin{equation}
 {\langle}0{\vert}b_{i}(z)\bar{b}_{j}(0){\vert}
 {\Upsilon}(p_{1},{\epsilon}_{\parallel}){\rangle}\,
 =\, \frac{1}{4}f_{\Upsilon}
 {\int}dk_{1}\,e^{-ik_{1}{\cdot}z}
  \Big\{ \!\!\not{\epsilon}_{\parallel} \Big[
   m_{1}\,{\phi}_{\Upsilon}^{v}(k_{1})
  -\!\!\not{p}_{1}\, {\phi}_{\Upsilon}^{t}(k_{1})
  \Big] \Big\}_{ji}
  \label{wave-bbl},
  \end{equation}
  \begin{equation}
 {\langle}B_{c}^{+}(p_{2}){\vert}\bar{c}_{i}(z)b_{j}(0){\vert}0{\rangle}\,
 =\, \frac{i}{4}f_{B_{c}} {\int}dk_{2}\,e^{ik_{2}{\cdot}z}\,
  \Big\{ {\gamma}_{5}\Big[ \!\!\not{p}_{2}\, {\phi}_{B_{c}}^{a}(k_{2})
  +m_{2}\,{\phi}_{B_{c}}^{p}(k_{2})\Big] \Big\}_{ji}
  \label{wave-bc},
  \end{equation}
  \begin{equation}
 {\langle}D_{q}^{-}(p_{3}){\vert}\bar{q}_{i}(z)c_{j}(0)
 {\vert}0{\rangle}\ =\
  \frac{i}{4}f_{D_{q}} {\int}_{0}^{1}dk_{3}\,e^{ik_{3}{\cdot}z}
  \Big\{ {\gamma}_{5}\Big[ \!\!\not{p}_{3}\,{\phi}_{D_{q}}^{a}(k_{3})
  +m_{3}\, {\phi}_{D_{q}}^{p}(k_{3}) \Big] \Big\}_{ji}
  \label{wave-ds},
  \end{equation}
  where $f_{\Upsilon}$, $f_{B_{c}}$, $f_{D_{q}}$ are
  decay constants.

  Because of the relations,
  $m_{{\Upsilon}(nS)}$ ${\simeq}$ $2m_{b}$,
  $m_{B_{c}}$ ${\simeq}$ $m_{b}$ $+$ $m_{c}$,
  and $m_{D_{q}}$ ${\simeq}$ $m_{c}$ $+$ $m_{q}$
  (see Table \ref{tab:input}),
  it might assume that the motion of the valence quarks
  in the considered mesons is nearly nonrelativistic.
  The wave functions of the ${\Upsilon}(nS)$, $B_{c}$, $D_{q}$
  mesons could be approximately described with the
  nonrelativistic quantum chromodynamics \cite{prd46,prd51,rmp77}
  and Schr\"{o}dinger equation.
  The wave functions of a nonrelativistic three-dimensional
  isotropic harmonic oscillator potential are given in
  Ref. \cite{plb751},
   \begin{equation}
  {\phi}_{{\Upsilon}(1S)}^{v}(x) = A\, x\bar{x}\,
  {\exp}\Big\{ -\frac{m_{b}^{2}}{8\,{\beta}_{1}^{2}\,x\,\bar{x}} \Big\}
   \label{wave-bbv},
   \end{equation}
   \begin{equation}
  {\phi}_{{\Upsilon}(1S)}^{t}(x) = B\, (x-\bar{x})^{2}\,
  {\exp}\Big\{ -\frac{m_{b}^{2}}{8\,{\beta}_{1}^{2}\,x\,\bar{x}} \Big\}
   \label{wave-bbt},
   \end{equation}
   \begin{equation}
  {\phi}_{{\Upsilon}(2S)}^{t,v}(x) = C\,
  {\phi}_{{\Upsilon}(1S)}^{t,v}(x)\,
   \Big\{ 1+\frac{m_{b}^{2}}{2\,{\beta}_{1}^{2}\,x\,\bar{x}} \Big\}
   \label{wave-bb2s},
   \end{equation}
   \begin{equation}
  {\phi}_{{\Upsilon}(3S)}^{t,v}(x) = D\,
  {\phi}_{{\Upsilon}(1S)}^{t,v}(x)\,
   \Big\{ \Big( 1-\frac{m_{b}^{2}}{2\,{\beta}_{1}^{2}\,x\,\bar{x}} \Big)^{2}
   +6 \Big\}
   \label{wave-bb3s},
   \end{equation}
   \begin{equation}
  {\phi}_{B_{c}}^{a}(x) = E\, x\bar{x}\,
  {\exp}\Big\{ -\frac{\bar{x}\,m_{c}^{2}+x\,m_{b}^{2}}
                     {8\,{\beta}_{2}^{2}\,x\,\bar{x}} \Big\}
   \label{wave-bca},
   \end{equation}
   \begin{equation}
  {\phi}_{B_{c}}^{p}(x) = F\,
  {\exp}\Big\{ -\frac{\bar{x}\,m_{c}^{2}+x\,m_{b}^{2}}
                     {8\,{\beta}_{2}^{2}\,x\,\bar{x}} \Big\}
   \label{wave-bcp},
   \end{equation}
   \begin{equation}
  {\phi}_{D_{q}}^{a}(x) = G\, x\bar{x}\,
  {\exp}\Big\{ -\frac{\bar{x}\,m_{q}^{2}+x\,m_{c}^{2}}
                     {8\,{\beta}_{3}^{2}\,x\,\bar{x}} \Big\}
   \label{wave-dsa},
   \end{equation}
   \begin{equation}
  {\phi}_{D_{q}}^{p}(x) = H\,
  {\exp}\Big\{ -\frac{\bar{x}\,m_{q}^{2}+x\,m_{c}^{2}}
                     {8\,{\beta}_{3}^{2}\,x\,\bar{x}} \Big\}
   \label{wave-dsp},
   \end{equation}
   where ${\beta}_{i}$ $=$ ${\xi}_{i}{\alpha}_{s}({\xi}_{i})$
   with ${\xi}_{i}$ $=$ $m_{i}/2$;
   parameters $A$, $B$, $C$, $D$, $E$, $F$, $G$, $H$ are the
   normalization coefficients satisfying the following conditions
   \begin{equation}
  {\int}_{0}^{1}dx\,{\phi}_{{\Upsilon}(nS)}^{v,t}(x) =1,
   \quad
  {\int}_{0}^{1}dx\,{\phi}_{B_{c}}^{a,p}(x)=1,
   \quad
  {\int}_{0}^{1}dx\,{\phi}_{D_{q}}^{a,p}(x)=1
   \label{wave-nom}.
   \end{equation}

  The shape lines of the distribution amplitudes
  ${\phi}_{{\Upsilon}(nS)}^{v,t}(x)$ and
  ${\phi}_{B_{c}}^{a,p}(x)$ have been displayed
  in Ref. \cite{plb751}, which are basically consistent with
  the physical picture that the valence quarks share momentums
  according to their masses.

  Here, one may question the nonrelativistic treatment on the
  wave functions of the $D_{s,d}$ mesons, because the motion
  of the light valence quark in $D$ meson is commonly
  assumed to be relativistic, and the behavior of the light
  valence quark in the heavy-light charmed $D_{s,d}$ mesons
  should be different from that in the heavy-heavy $B_{c}$
  and ${\Upsilon}(nS)$ mesons.
  In addition, there are several phenomenological models for
  the $D_{s,d}$ meson wave functions, for example, Eq.(30)
  in Ref. \cite{prd78lv}.
  The $D$ wave function, which is widely used within the pQCD
  framework, and is also favored by Ref. \cite{prd78lv} via
  fitting with measurements on the $B$ ${\to}$ $DP$ decays,
  is written as
   \begin{equation}
  {\phi}_{D}(x,b) = 6\,x\bar{x}\,\Big\{1+C_{D}(1-2x)\Big\}
  {\exp}\Big\{ -\frac{1}{2}w^{2}b^{2} \Big\}
   \label{wave-d-lv},
   \end{equation}
  where $C_{D}$ $=$ $0.5$ and $w$ $=$ $0.1$ GeV for the $D_{d}$ meson;
  $C_{D}$ $=$ $0.4$ and $w$ $=$ $0.2$ GeV for the $D_{s}$ meson;
  the exponential function represents the $k_{T}$ distribution.
  The same model of Eq.(\ref{wave-d-lv}) is usually taken as the
  twist-2 and twist-3 distribution amplitudes in many practical
  applications \cite{prd78lv}.

  \begin{figure}[h]
  \includegraphics[width=0.98\textwidth,bb=80 610 530 720]{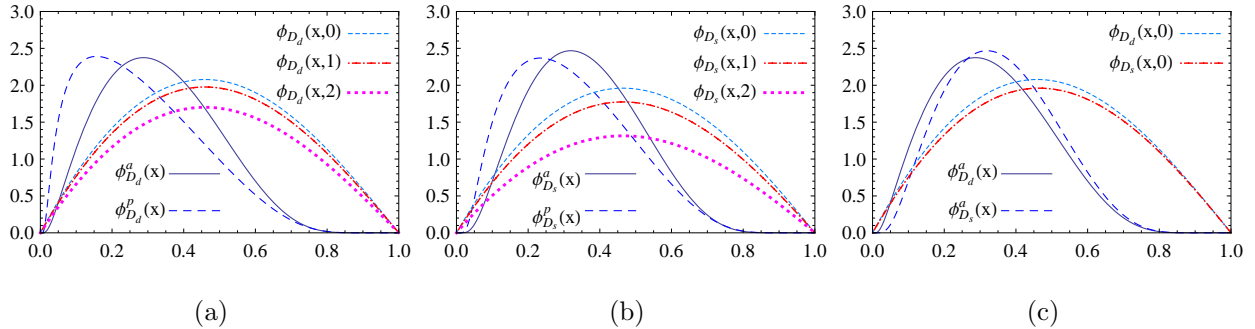}
  \caption{The distributions of the $D_{d}$ meson wave functions in (a),
  the distributions of the $D_{s}$ meson wave functions in (b), and
  the distributions of the $D_{s,d}$ meson wave functions in (c),
  where ${\phi}_{D_{q}}^{a}(x)$, ${\phi}_{D_{q}}^{p}(x)$, and
  ${\phi}_{D}(x,b)$ correspond to the Eq.(\ref{wave-dsa}),
  Eq.(\ref{wave-dsp}), and Eq.(\ref{wave-d-lv}), respectively.}
  \label{fig:wave}
  \end{figure}

  To show that the nonrelativistic description of the $D_{s,d}$
  wave functions seems to be acceptable, the shape lines of the
  $D$ wave functions are displayed in Fig.\ref{fig:wave}.
  It is clearly seen from Fig.\ref{fig:wave} that the shape lines
  of both Eq.(\ref{wave-dsa}) and Eq.(\ref{wave-dsp}) have a broad
  peak at the small $x$ regions, while the distributions of
  Eq.(\ref{wave-d-lv}) is nearly symmetric to the variable $x$.
  This fact may imply that although the nonrelativistic model of
  the $D$ wave functions is crude, Eq.(\ref{wave-dsa}) and
  Eq.(\ref{wave-dsp}) can reflect, at least to some extent,
  the feature that the light valence quark might carry less
  momentums than the charm quark in the $D_{s,d}$
  mesons. In addition, the flavor asymmetric effects, and
  the difference between the twist-2 and twist-3 distribution
  amplitudes are considered at least in part by
  Eq.(\ref{wave-dsa}) and Eq.(\ref{wave-dsp}).
  In the following calculation, we will use Eq.(\ref{wave-dsa}) and
  Eq.(\ref{wave-dsp}) as the twist-2 and twist-3 distribution
  amplitudes of the $D_{s,d}$ meson, respectively.

  \subsection{Decay amplitudes}
  \label{sec0205}
\begin{figure}[h]
  \includegraphics[width=0.95\textwidth,bb=80 530 530 720]{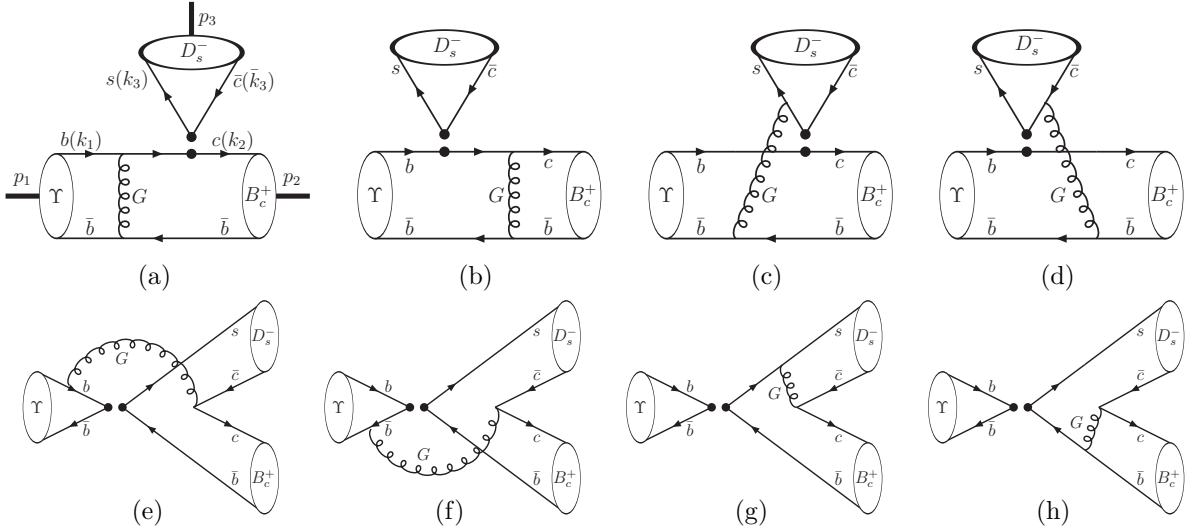}
  \caption{Feynman diagrams for the ${\Upsilon}(nS)$ ${\to}$
  $B_{c}D_{s}$ decay with the pQCD approach, including
  the factorizable emission diagrams (a,b),
  the nonfactorizable emission diagrams (c,d),
  the nonfactorizable annihilation diagrams (e,f),
  and the factorizable annihilation diagrams (g,h).}
  \label{fig:fey}
  \end{figure}

  The Feynman diagrams for the ${\Upsilon}(nS)$ ${\to}$
  $B_{c}D_{s}$ decay are shown in Fig.\ref{fig:fey}.
  There are two types: the emission and annihilation
  topologies, where diagram with gluon attaching to quarks
  in the same meson and between two different mesons
  are entitled factorizable and nonfactorizable diagrams,
  respectively.

  By calculating these diagrams with the pQCD master
  formula Eq.(\ref{hadronic}), the decay amplitudes
  of ${\Upsilon}(nS)$ ${\to}$ $B_{c}D_{q}$
  decays (where $q$ $=$ $d$, $s$) can be expressed as:
   \begin{eqnarray}
   \lefteqn{ {\cal A}({\Upsilon}(nS){\to}B_{c}D_{q})\ =\
   \sqrt{2}\,G_{F}\,{\pi}\,f_{\Upsilon}\,f_{B_{c}}\,f_{D_{q}}\,
   \frac{C_{F}}{N}\,m_{{\Upsilon}}^{3}\,({\epsilon}_{\Upsilon}{\cdot}p_{D_{q}}) }
   \nonumber \\ &{\times}&
   \Big\{ V_{cb} V_{cq}^{\ast}\,\Big[ {\cal A}_{a+b}^{LL}\,a_{1}
   +{\cal A}_{c+d}^{LL}\,C_{2} \Big]
   - V_{tb} V_{tq}^{\ast}\,\Big[ {\cal A}_{a+b}^{LL}\,(a_{4}+a_{10})
   \nonumber \\ & &
   + {\cal A}_{a+b}^{SP}\, (a_{6}+a_{8})
   + {\cal A}_{c+d}^{LL}\, (C_{3}+C_{9})
   + {\cal A}_{c+d}^{SP}\, (C_{5}+C_{7})
   \nonumber \\ & &
   + {\cal A}_{e+f}^{LL}\, (C_{3}+C_{4}-\frac{1}{2}C_{9}-\frac{1}{2}C_{10})
   + {\cal A}_{e+f}^{LR}\, (C_{6}-\frac{1}{2}C_{8})
   \nonumber \\ & &
   + {\cal A}_{g+h}^{LL}\, (a_{3}+a_{4}-\frac{1}{2}a_{9}-\frac{1}{2}a_{10})
   + {\cal A}_{g+h}^{LR}\, (a_{5}-\frac{1}{2}a_{7})
   \nonumber \\ & &
   + {\cal A}_{e+f}^{SP}\, (C_{5}-\frac{1}{2}C_{7}) \Big] \Big\}
   \label{amp-all},
   \end{eqnarray}
  where $C_{F}$ $=$ $4/3$ and the color number $N$ $=$ $3$.

  The parameters $a_{i}$ are defined as follows.
  \begin{eqnarray}
  a_{i} &=& C_{i}+C_{i+1}/N, \quad (i=1,3,5,7,9);
  \label{eq:ai1357} \\
  a_{i} &=& C_{i}+C_{i-1}/N, \quad (i=2,4,5,6,10).
  \label{eq:ai2468}
  \end{eqnarray}

  The building blocks ${\cal A}_{a+b}$, ${\cal A}_{c+d}$,
  ${\cal A}_{e+f}$, ${\cal A}_{g+h}$ denote the contributions of
  the factorizable emission diagrams Fig.\ref{fig:fey}(a,b),
  the nonfactorizable emission diagrams Fig.\ref{fig:fey}(c,d),
  the nonfactorizable annihilation diagrams Fig.\ref{fig:fey}(e,f),
  the factorizable annihilation diagrams Fig.\ref{fig:fey}(g,h),
  respectively. They are defined as
  \begin{equation}
  {\cal A}_{i+j}^{k} = {\cal A}_{i}^{k}+{\cal A}_{j}^{k}
  \label{eq:ampij},
  \end{equation}
  where the subscripts $i$ and $j$ correspond
  to the indices of Fig.\ref{fig:fey};
  the superscript $k$ refers
  to one of the three possible Dirac structures, namely
  $k$ $=$ $LL$ for $(V-A){\otimes}(V-A)$,
  $k$ $=$ $LR$ for $(V-A){\otimes}(V+A)$, and
  $k$ $=$ $SP$ for $-2(S-P){\otimes}(S+P)$.
  The explicit expressions of these building blocks
  are collected in the Appendix \ref{blocks}.

  \section{Numerical results and discussion}
  \label{sec03}
  In the rest frame of the ${\Upsilon}(nS)$ meson,
  the $CP$-averaged branching ratios for the ${\Upsilon}(nS)$
  ${\to}$ $B_{c}D_{s,d}$ weak decays are written as
   \begin{equation}
  {\cal B}r({\Upsilon}(nS){\to}B_{c}D_{s,d})\ =\ \frac{1}{12{\pi}}\,
   \frac{p}{m_{{\Upsilon}}^{2}{\Gamma}_{{\Upsilon}}}\,
  {\vert}{\cal A}({\Upsilon}(nS){\to}B_{c}D_{s,d}){\vert}^{2}
   \label{br}.
   \end{equation}

   \begin{table}[h]
   \caption{The numerical values of some input parameters.}
   \label{tab:input}
   \begin{ruledtabular}
   \begin{tabular}{ll}
   \multicolumn{2}{c}{The Wolfenstein parameters\footnotemark[2]} \\ \hline
     $A$ $=$ $0.814^{+0.023}_{-0.024}$ \cite{pdg},
   & ${\lambda}$ $=$ $0.22537{\pm}0.00061$ \cite{pdg}, \\
     $\bar{\rho}$ $=$ $0.117{\pm}0.021$ \cite{pdg},
   & $\bar{\eta}$ $=$ $0.353{\pm}0.013$ \cite{pdg}, \\ \hline
   \multicolumn{2}{c}{Mass and decay constant} \\ \hline
     $m_{b}$ $=$ $4.78{\pm}0.06$ GeV \cite{pdg},
   & $m_{c}$ $=$ $1.67{\pm}0.07$ GeV \cite{pdg}, \\
     $m_{s}$ ${\simeq}$ $510$ MeV \cite{book},
   & $m_{d}$ ${\simeq}$ $310$ MeV \cite{book}, \\
     $m_{B_{c}}$ $=$ $6275.6{\pm}1.1$ MeV \cite{pdg},
   & $m_{D_{s}}$ $=$ $1968.30{\pm}0.11$ MeV \cite{pdg}, \\
     $m_{D_{d}}$ $=$ $1869.61{\pm}0.10$ MeV \cite{pdg}, & \\ \hline
     $f_{{\Upsilon}(1S)}$ $=$ $676.4{\pm}10.7$ MeV \cite{plb751}
   & $f_{B_{c}}$ $=$ $489{\pm}5$ MeV \cite{fbc}, \\
     $f_{{\Upsilon}(2S)}$ $=$ $473.0{\pm}23.7$ MeV \cite{plb751}
   & $f_{D_{s}}$ $=$ $257.5{\pm}4.6$ MeV \cite{pdg}, \\
     $f_{{\Upsilon}(3S)}$ $=$ $409.5{\pm}29.4$ MeV \cite{plb751}
   & $f_{D_{d}}$ $=$ $204.6{\pm}5.0$ MeV \cite{pdg}.
   \end{tabular}
   \end{ruledtabular}
   \footnotetext[2]{The relation between parameters (${\rho}$, ${\eta}$)
   and ($\bar{\rho}$, $\bar{\eta}$) is \cite{pdg}: $({\rho}+i{\eta})$ $=$
    $\displaystyle \frac{ \sqrt{1-A^{2}{\lambda}^{4}}(\bar{\rho}+i\bar{\eta}) }
    { \sqrt{1-{\lambda}^{2}}[1-A^{2}{\lambda}^{4}(\bar{\rho}+i\bar{\eta})] }$.}
   \end{table}

   \begin{table}[h]
   \caption{The $CP$-averaged branching ratios for the
    ${\Upsilon}(nS)$ ${\to}$ $B_{c}D_{s,d}$ weak decays.}
   \label{tab:br}
   \begin{ruledtabular}
   \begin{tabular}{cc}
   decay mode & ${\cal B}r$ \\ \hline
    ${\Upsilon}(1S)$ ${\to}$ $B_{c}D_{s}$
  & $(5.42^{+0.38+0.64+1.47}_{-0.37-0.60-0.76}){\times}10^{-10}$ \\
    ${\Upsilon}(2S)$ ${\to}$ $B_{c}D_{s}$
  & $(4.28^{+0.30+0.49+0.93}_{-0.29-0.67-0.48}){\times}10^{-10}$ \\
    ${\Upsilon}(3S)$ ${\to}$ $B_{c}D_{s}$
  & $(4.61^{+0.33+0.40+0.93}_{-0.31-0.88-0.52}){\times}10^{-10}$ \\ \hline
    ${\Upsilon}(1S)$ ${\to}$ $B_{c}D_{d}$
  & $(1.96^{+0.15+0.23+0.56}_{-0.15-0.22-0.27}){\times}10^{-11}$ \\
    ${\Upsilon}(2S)$ ${\to}$ $B_{c}D_{d}$
  & $(1.38^{+0.11+0.24+0.29}_{-0.10-0.05-0.15}){\times}10^{-11}$ \\
    ${\Upsilon}(3S)$ ${\to}$ $B_{c}D_{d}$
  & $(1.58^{+0.12+0.15+0.33}_{-0.12-0.23-0.16}){\times}10^{-11}$
   \end{tabular}
   \end{ruledtabular}
   \end{table}

  The input parameters are listed in Table \ref{tab:bb}
  and \ref{tab:input}. If not specified explicitly, we
  will take their central values as the default inputs.
  The numerical results on the $CP$-averaged branching
  ratios for the ${\Upsilon}(nS)$ ${\to}$ $B_{c}D_{s,d}$
  weak decays are listed in Table \ref{tab:br}, where
  the first uncertainties come from the CKM parameters;
  the second uncertainties are due to the variation of
  mass $m_{b}$ and $m_{c}$;
  the third uncertainties arise from the typical scale
  ${\mu}$ $=$ $(1{\pm}0.1)t_{i}$ and the expressions
  of $t_{i}$ for different topologies are given in
  Eqs.(\ref{tab}-\ref{tgh}).
  The following are some comments.

  (1) Because of the relation between the CKM factors
  ${\vert}V_{cb}V_{cs}^{\ast}{\vert}$ $>$
  ${\vert}V_{cb}V_{cd}^{\ast}{\vert}$,
  and the relation between decay constants
  $f_{D_{s}}$ $>$ $f_{D_{d}}$,
  there is a hierarchical relation between branching ratios,
  i.e., ${\cal B}r({\Upsilon}(nS){\to}B_{c}D_{s})$ $>$
  ${\cal B}r({\Upsilon}(nS){\to}B_{c}D_{d})$ for the
  same quantum number $n$.

  (2) The relation among mass
  $m_{{\Upsilon}(3S)}$ $>$ $m_{{\Upsilon}(2S)}$ $>$ $m_{{\Upsilon}(1S)}$
  and total width ${\Gamma}_{{\Upsilon}(3S)}$ $<$ ${\Gamma}_{{\Upsilon}(2S)}$
  $<$ ${\Gamma}_{{\Upsilon}(1S)}$ should in principle result
  in the relation among branching ratios,
  ${\cal B}r({\Upsilon}(3S){\to}B_{c}D_{q})$ $>$
  ${\cal B}r({\Upsilon}(2S){\to}B_{c}D_{q})$ $>$
  ${\cal B}r({\Upsilon}(1S){\to}B_{c}D_{q})$.
  The numbers in Table \ref{tab:br} show that
  branching ratios for the ${\Upsilon}(nS)$ ${\to}$
  $B_{c}D_{q})$ weak decays seem to be close to each other,
  and have almost nothing with the radial quantum number $n$.
  The reason may be that the decay amplitudes are
  proportional to decay constant $f_{{\Upsilon}(nS)}$,
  and hence there is an approximation,
   \begin{eqnarray}
   & &
   {\cal B}r({\Upsilon}(1S){\to}B_{c}D_{q}) :
   {\cal B}r({\Upsilon}(2S){\to}B_{c}D_{q}) :
   {\cal B}r({\Upsilon}(3S){\to}B_{c}D_{q})
   \nonumber \\ &{\propto}&
   \frac{ f_{{\Upsilon}(1S)}^{2} }{ {\Gamma}_{{\Upsilon}(1S)} } :
   \frac{ f_{{\Upsilon}(2S)}^{2} }{ {\Gamma}_{{\Upsilon}(2S)} } :
   \frac{ f_{{\Upsilon}(3S)}^{2} }{ {\Gamma}_{{\Upsilon}(3S)} }\
   {\simeq}\
   1 : 1 : 1
   \label{f2g}.
   \end{eqnarray}

  (3) Although different wave functions are used for the
  $D_{s,d}$ meson in the calculation due to the mass relation
  $m_{s}$ ${\neq}$ $m_{d}$ and $m_{D_{s}}$ ${\neq}$ $m_{D_{d}}$,
  the flavor symmetry breaking effects mainly appear in
  the CKM parameters and the decay constant $f_{D_{s,d}}$,
   \begin{equation}
   \frac{ {\cal B}r({\Upsilon}(nS){\to}B_{c}D_{s}) }
        { {\cal B}r({\Upsilon}(nS){\to}B_{c}D_{d}) }\
  {\simeq}\
   \frac{ {\vert}V_{cb}V_{cs}^{\ast}{\vert}^{2}\, f_{D_{s}}^{2} }
        { {\vert}V_{cb}V_{cd}^{\ast}{\vert}^{2}\, f_{D_{d}}^{2} }
   \label{eq:uspin},
   \end{equation}
   for the same radial quantum number $n$.

  (4) Compared the ${\Upsilon}(nS)$ ${\to}$ $B_{c}D_{s}$ decay
  with the ${\Upsilon}(nS)$ ${\to}$ $B_{c}{\pi}$ decay \cite{plb751},
  they are both color-favored and CKM-favored.
  Only the emission topologies, and only the tree operators,
  contribute to the ${\Upsilon}(nS)$ ${\to}$ $B_{c}{\pi}$ decay,
  while both emission and annihilation topologies, and
  both tree and penguin operators, contribute to the
  ${\Upsilon}(nS)$ ${\to}$ $B_{c}D_{s}$ decay.
  In addition, the penguin contributions are dynamically
  enhanced due to the typical scale $t$ within
  the pQCD framework \cite{plb504}.
  These might explain the fact that although the final phase
  spaces for the ${\Upsilon}(nS)$ ${\to}$ $B_{c}D_{s}$ decay
  are more compact than those for the ${\Upsilon}(nS)$
  ${\to}$ $B_{c}{\pi}$ decay, there is still the
  relation\footnotemark[2]
  \footnotetext[2]{The branching ratio for the
  ${\Upsilon}(nS)$ ${\to}$ $B_{c}{\pi}$ decay is
  about ${\cal O}(10^{-11})$ \cite{plb751}
  with the pQCD approach.}
  between branching ratios
  ${\cal B}r({\Upsilon}(nS){\to}B_{c}D_{s})$
  $>$ ${\cal B}r({\Upsilon}(nS){\to}B_{c}{\pi})$.

  (5)
  It is seen that branching ratios for
  the ${\Upsilon}(nS)$ ${\to}$ $B_{c}D_{s}$ ($B_{c}D_{d}$)
  decay can reach up to $10^{-10}$ ($10^{-11}$), which
  might be accessible at the running LHC and forthcoming
  SuperKEKB.
  For example, the ${\Upsilon}(nS)$ production cross
  section in p-Pb collision is a few ${\mu}b$
  with the LHCb \cite{jhep1407} and ALICE \cite{plb740}
  detectors at LHC.
  Over $10^{12}$ ${\Upsilon}(nS)$ mesons per $ab^{-1}$
  data collected at LHCb and ALICE are in principle available,
  corresponding to a few hundreds (tens) of the ${\Upsilon}(nS)$
  ${\to}$ $B_{c}D_{s}$ ($B_{c}D_{d}$) events.

  (6) Besides the uncertainties listed in Table \ref{tab:br},
  the decay constants can bring about 5\%, 10\%, 15\%
  uncertainties to branching ratios for the ${\Upsilon}(1S)$
  ${\Upsilon}(2S)$, ${\Upsilon}(3S)$ mesons decay into
  the $B_{c}D_{s,d}$ states, respectively, mainly from
  $f_{{\Upsilon}(2S,3S)}$.
  Other factors, such as the contributions of higher
  order corrections to HME, relativistic effects, different
  models for the wave functions, and so on, deserve
  the dedicated study. Our results just provide an order
  of magnitude estimation.

  \section{Summary}
  \label{sec04}
  The ${\Upsilon}(nS)$ weak decay is allowable within the
  standard model, although the branching ratio is tiny and
  the experimental search is very difficult.
  With the potential prospects of the ${\Upsilon}(nS)$
  at high-luminosity dedicated heavy-flavor factories,
  the ${\Upsilon}(nS)$ ${\to}$ $B_{c}D_{s,d}$
  weak decays are studied with the pQCD approach firstly.
  It is found that with the nonrelativistic wave functions
  for ${\Upsilon}(nS)$, $B_{c}$, and $D_{s,d}$ mesons,
  branching ratios ${\cal B}r({\Upsilon}(nS){\to}B_{c}D_{s})$
  ${\sim}$ ${\cal O}(10^{-10})$ and
  ${\cal B}r({\Upsilon}(nS){\to}B_{c}D_{d})$
  ${\sim}$ ${\cal O}(10^{-11})$, which might be measurable in
  the future experiments.

  \section*{Acknowledgments}
  We thank Professor Dongsheng Du (IHEP@CAS) and
  Professor Yadong Yang (CCNU) for helpful discussion.

  \begin{appendix}
  \section{The building blocks of decay amplitudes}
  \label{blocks}
  For the sake of simplicity,
  we decompose the decay amplitude Eq.(\ref{amp-all})
  into some building blocks ${\cal A}_{i}^{k}$, where
  the subscript $i$ on ${\cal A}_{i}^{k}$ corresponds to
  the indices of Fig.\ref{fig:fey};
  the superscript $k$ on ${\cal A}_{i}^{k}$ refers
  to one of the three possible Dirac structures
  ${\Gamma}_{1}{\otimes}{\Gamma}_{2}$ of the
  four-quark operator
  $(\bar{q}_{1}{\Gamma}_{1}q_{2})(\bar{q}_{1}{\Gamma}_{2}q_{2})$,
  namely
  $k$ $=$ $LL$ for $(V-A){\otimes}(V-A)$,
  $k$ $=$ $LR$ for $(V-A){\otimes}(V+A)$, and
  $k$ $=$ $SP$ for $-2(S-P){\otimes}(S+P)$.
  The explicit expressions of ${\cal A}_{i}^{k}$
  are written as follows.
   \begin{eqnarray}
  {\cal A}_{a}^{LL} &=&
  {\int}_{0}^{1}dx_{1} {\int}_{0}^{1}dx_{2}
  {\int}_{0}^{\infty}b_{1} db_{1} {\int}_{0}^{\infty}b_{2} db_{2}\,
  H_{a}({\alpha}_{e},{\beta}_{a},b_{1},b_{2})\, E_{a}(t_{a})
   \nonumber \\ & &
  {\alpha}_{s}(t_{a})\, {\phi}_{\Upsilon}^{v}(x_{1})\,  \Big\{
  {\phi}_{B_{c}}^{a}(x_{2})\Big[ x_{2}+r_{3}^{2}\,\bar{x}_{2} \Big]
 +{\phi}_{B_{c}}^{p}(x_{2})\, r_{2}\,r_{b} \Big\}
   \label{amp-figa-01}, \\
  {\cal A}_{a}^{SP} &=&
   -2\,r_{3}\,{\int}_{0}^{1}dx_{1} {\int}_{0}^{1}dx_{2}
   {\int}_{0}^{\infty}b_{1} db_{1} {\int}_{0}^{\infty}b_{2} db_{2}\,
  H_{a}({\alpha}_{e},{\beta}_{a},b_{1}, b_{2})
   \nonumber \\ & &
  E_{a}(t_{a})\, {\alpha}_{s}(t_{a})\,
  {\phi}_{\Upsilon}^{v}(x_{1})\, \Big\{
  {\phi}_{B_{c}}^{a}(x_{2})\,r_{b}
 +{\phi}_{B_{c}}^{p}(x_{2})\,r_{2}\,\bar{x}_{2} \Big\}
   \label{amp-figa-03},
   \end{eqnarray}
   \begin{eqnarray}
  {\cal A}_{b}^{LL} &=&
  {\int}_{0}^{1}dx_{1} {\int}_{0}^{1}dx_{2}
  {\int}_{0}^{\infty}b_{1} db_{1} {\int}_{0}^{\infty}b_{2} db_{2}\,
  H_{b}({\alpha}_{e},{\beta}_{b},b_{2},b_{1})\, E_{b}(t_{b})
   \nonumber \\ & &
  {\alpha}_{s}(t_{b})\, \Big\{ {\phi}_{\Upsilon}^{v}(x_{1}) \Big[
  2\,{\phi}_{B_{c}}^{p}(x_{2})\,r_{2}\,r_{c}
 -{\phi}_{B_{c}}^{a}(x_{2})\, (r_{2}^{2}\,x_{1}+r_{3}^{2}\,\bar{x}_{1}) \Big]
   \nonumber \\ & &
   + {\phi}_{\Upsilon}^{t}(x_{1}) \Big[
  2\,{\phi}_{B_{c}}^{p}(x_{2})\, r_{2}\,x_{1}
 -{\phi}_{B_{c}}^{a}(x_{2})\,r_{c} \Big] \Big\}
   \label{amp-figb-01}, \\
  {\cal A}_{b}^{SP} &=& -2\,r_{3}\,
  {\int}_{0}^{1}dx_{1} {\int}_{0}^{1}dx_{2}
  {\int}_{0}^{\infty}b_{1} db_{1} {\int}_{0}^{\infty}b_{2} db_{2}\,
  H_{b}({\alpha}_{e},{\beta}_{b},b_{2},b_{1})
   \nonumber \\ & &
   E_{b}(t_{b})\, {\alpha}_{s}(t_{b})\,
   \Big\{ {\phi}_{\Upsilon}^{v}(x_{1}) \Big[
  2\,{\phi}_{B_{c}}^{p}(x_{2})\, r_{2}
 -{\phi}_{B_{c}}^{a}(x_{2})\,r_{c} \Big]
   \nonumber \\ & &
 -{\phi}_{\Upsilon}^{t}(x_{1})\,{\phi}_{B_{c}}^{a}(x_{2})\,
   \bar{x}_{1}\Big\}
   \label{amp-figb-03},
   \end{eqnarray}
   \begin{eqnarray}
  {\cal A}_{c}^{LL} &=& \frac{1}{N}
  {\int}_{0}^{1}dx_{1} {\int}_{0}^{1}dx_{2} {\int}_{0}^{1}dx_{3}
  {\int}_{0}^{\infty}db_{1} {\int}_{0}^{\infty}b_{2} db_{2}
  {\int}_{0}^{\infty}b_{3} db_{3}\,{\delta}(b_{1}-b_{2})\,
  {\alpha}_{s}(t_{c})
   \nonumber \\ & &
  H_{cd}({\alpha}_{e},{\beta}_{c},b_{2},b_{3})\,
  E_{c}(t_{c})\, {\phi}_{D_{q}}^{a}(x_{3})\,
   \Big\{ {\phi}_{\Upsilon}^{t}(x_{1})\,
  {\phi}_{B_{c}}^{p}(x_{2})\, r_{2}\,(x_{2}-x_{1})
   \nonumber \\ & &
 +{\phi}_{\Upsilon}^{v}(x_{1})\,{\phi}_{B_{c}}^{a}(x_{2})\,
   \Big[ \frac{s\,(x_{1}-\bar{x}_{3}) }{m_{1}^{2}}\,
   +2\,r_{2}^{2}\,(x_{1}-x_{2}) \Big] \Big\}
   \label{amp-figc-01}, \\
  {\cal A}_{c}^{SP} &=& -\frac{1}{N}\,r_{3}
  {\int}_{0}^{1}dx_{1} {\int}_{0}^{1}dx_{2} {\int}_{0}^{1}dx_{3}
  {\int}_{0}^{\infty} db_{1} {\int}_{0}^{\infty}b_{2} db_{2}
  {\int}_{0}^{\infty}b_{3} db_{3}\,{\delta}(b_{1}-b_{2})
   \nonumber \\ & &
  H_{cd}({\alpha}_{e},{\beta}_{c},b_{2},b_{3})\,
  E_{c}(t_{c})\,{\alpha}_{s}(t_{c})\, {\phi}_{D_{q}}^{p}(x_{3})\,
   \Big\{ {\phi}_{\Upsilon}^{t}(x_{1})\,{\phi}_{B_{c}}^{a}(x_{2})\,
    (x_{1}-\bar{x}_{3})
   \nonumber \\ & &
 +{\phi}_{\Upsilon}^{v}(x_{1})\,{\phi}_{B_{c}}^{p}(x_{2})\,
   r_{2}\, (\bar{x}_{3}-x_{2})  \Big\}
   \label{amp-figc-03},
   \end{eqnarray}
   \begin{eqnarray}
  {\cal A}_{d}^{LL} &=& \frac{1}{N}
  {\int}_{0}^{1}dx_{1} {\int}_{0}^{1}dx_{2} {\int}_{0}^{1}dx_{3}
  {\int}_{0}^{\infty}db_{1} {\int}_{0}^{\infty}b_{2} db_{2}
  {\int}_{0}^{\infty}b_{3} db_{3}\,{\delta}(b_{1}-b_{2})\,
  {\alpha}_{s}(t_{d})\,
   \nonumber \\ & &
  H_{cd}({\alpha}_{e},{\beta}_{d},b_{2},b_{3})\, E_{d}(t_{d})\,
   \Big\{ {\phi}_{\Upsilon}^{t}(x_{1})\,{\phi}_{B_{c}}^{p}(x_{2})\,
  {\phi}_{D_{q}}^{a}(x_{3})\,r_{2}\,(x_{2}-x_{1})
   \nonumber \\ & &
 +{\phi}_{\Upsilon}^{v}(x_{1})\,{\phi}_{B_{c}}^{a}(x_{2})\,
  \Big[ {\phi}_{D_{q}}^{a}(x_{3})\,\frac{s\,(x_{3}-x_{2})}{m_{1}^{2}}
  -{\phi}_{D_{q}}^{p}(x_{3})\,r_{3}\,r_{c} \Big] \Big\}
   \label{amp-figd-01}, \\
  {\cal A}_{d}^{SP} &=& -\frac{1}{N}\,r_{3}
  {\int}_{0}^{1}dx_{1} {\int}_{0}^{1}dx_{2} {\int}_{0}^{1}dx_{3}
  {\int}_{0}^{\infty}db_{1} {\int}_{0}^{\infty}b_{2} db_{2}
  {\int}_{0}^{\infty}b_{3} db_{3}\
  \nonumber \\ & &
  {\delta}(b_{1}-b_{2})\,
  H_{cd}({\alpha}_{e},{\beta}_{d},b_{2},b_{3})\,
  E_{d}(t_{d})\,{\alpha}_{s}(t_{d})\,
   \nonumber \\ & &
  \Big\{ {\phi}_{\Upsilon}^{v}(x_{1})\, {\phi}_{B_{c}}^{p}(x_{2})\,r_{2}\,
  \Big[ {\phi}_{D_{q}}^{a}(x_{3})\, r_{c}/r_{3}
  +{\phi}_{D_{q}}^{p}(x_{3})\,(x_{2}-x_{3}) \Big]
  \nonumber \\ & &
 +{\phi}_{\Upsilon}^{t}(x_{1})\,{\phi}_{B_{c}}^{a}(x_{2})\,
  \Big[ {\phi}_{D_{q}}^{p}(x_{3})\,(x_{3}-x_{1})
  -{\phi}_{D_{q}}^{a}(x_{3})\,r_{c}/r_{3} \Big] \Big\}
   \label{amp-figd-03},
   \end{eqnarray}
   \begin{eqnarray}
  {\cal A}_{e}^{LL} &=& \frac{1}{N}
  {\int}_{0}^{1}dx_{1} {\int}_{0}^{1}dx_{2} {\int}_{0}^{1}dx_{3}
  {\int}_{0}^{\infty}b_{1} db_{1} {\int}_{0}^{\infty}b_{2} db_{2}
  {\int}_{0}^{\infty} db_{3}\, {\delta}(b_{2}-b_{3})\, {\alpha}_{s}(t_{e})
   \nonumber \\ & &
  H_{ef}({\alpha}_{a},{\beta}_{e},b_{1},b_{2})\, E_{e}(t_{e})\,
   \Big\{ {\phi}_{\Upsilon}^{v}(x_{1})
   \Big[ {\phi}_{B_{c}}^{p}(x_{2})\, {\phi}_{D_{q}}^{p}(x_{3})\,
   r_{2}\,r_{3}\,(x_{2}-\bar{x}_{3})
   \nonumber \\ & &
 +{\phi}_{B_{c}}^{a}(x_{2})\, {\phi}_{D_{q}}^{a}(x_{3})\,
   \big\{ \frac{s\,(x_{1}-\bar{x}_{3})}{m_{1}^{2}}
   +2\,r_{2}^{2}\,(x_{1}-x_{2}) \big\} \Big]
   \nonumber \\ & &
   -r_{b}\,{\phi}_{\Upsilon}^{t}(x_{1})\,
   {\phi}_{B_{c}}^{a}(x_{2})\, {\phi}_{D_{q}}^{a}(x_{3}) \Big\}
   \label{amp-fige-01}, \\
  {\cal A}_{e}^{LR} &=& \frac{1}{N}
  {\int}_{0}^{1}dx_{1} {\int}_{0}^{1}dx_{2} {\int}_{0}^{1}dx_{3}
  {\int}_{0}^{\infty}b_{1} db_{1} {\int}_{0}^{\infty}b_{2} db_{2}
  {\int}_{0}^{\infty} db_{3}\, {\delta}(b_{2}-b_{3})\, {\alpha}_{s}(t_{e})
   \nonumber \\ & &
   H_{ef}({\alpha}_{a},{\beta}_{e},b_{1},b_{2})\, E_{e}(t_{e})\,
   \Big\{ {\phi}_{\Upsilon}^{v}(x_{1}) \Big[
  {\phi}_{B_{c}}^{p}(x_{2})\, {\phi}_{D_{q}}^{p}(x_{3})\,
   r_{2}\,r_{3}\, (x_{2}-\bar{x}_{3})
   \nonumber \\ & &
 +{\phi}_{B_{c}}^{a}(x_{2})\, {\phi}_{D_{q}}^{a}(x_{3})\,
   \big\{ \frac{s\,(x_{2}-x_{1})}{m_{1}^{2}}
   +2\,r_{3}^{2}\, (\bar{x}_{3}-x_{1}) \big\}
   \nonumber \\ & &
   +r_{b}\,{\phi}_{\Upsilon}^{t}(x_{1})\,
   {\phi}_{B_{c}}^{a}(x_{2})\, {\phi}_{D_{q}}^{a}(x_{3}) \Big\}
   \label{amp-fige-02}, \\
  {\cal A}_{e}^{SP} &=& \frac{1}{N}
  {\int}_{0}^{1}dx_{1} {\int}_{0}^{1}dx_{2} {\int}_{0}^{1}dx_{3}
  {\int}_{0}^{\infty}b_{1} db_{1} {\int}_{0}^{\infty}b_{2} db_{2}
  {\int}_{0}^{\infty} db_{3}\, {\delta}(b_{2}-b_{3})\,{\alpha}_{s}(t_{e})
   \nonumber \\ & &
  H_{ef}({\alpha}_{a},{\beta}_{e},b_{1},b_{2})\,E_{e}(t_{e})\,
   \Big\{ {\phi}_{\Upsilon}^{t}(x_{1}) \Big[
  {\phi}_{B_{c}}^{a}(x_{2})\, {\phi}_{D_{q}}^{p}(x_{3})\,r_{3}\,
  (\bar{x}_{3}-x_{1})
   \nonumber \\ & &
 +{\phi}_{B_{c}}^{p}(x_{2})\, {\phi}_{D_{q}}^{a}(x_{3})\,r_{2}\,
  (x_{2}-x_{1}) \Big]
 +{\phi}_{\Upsilon}^{v}(x_{1})\,r_{b} \Big[
  {\phi}_{B_{c}}^{a}(x_{2})\, {\phi}_{D_{q}}^{p}(x_{3})\,r_{3}
   \nonumber \\ & &
 +{\phi}_{B_{c}}^{p}(x_{2})\, {\phi}_{D_{q}}^{a}(x_{3})\,r_{2} \Big] \Big\}
   \label{amp-fige-03},
   \end{eqnarray}
   \begin{eqnarray}
  {\cal A}_{f}^{LL} &=& \frac{1}{N}
  {\int}_{0}^{1}dx_{1} {\int}_{0}^{1}dx_{2} {\int}_{0}^{1}dx_{3}
  {\int}_{0}^{\infty}b_{1} db_{1} {\int}_{0}^{\infty}b_{2} db_{2}
  {\int}_{0}^{\infty} db_{3}\, {\delta}(b_{2}-b_{3})\,{\alpha}_{s}(t_{f})
   \nonumber \\ & &
  H_{ef}({\alpha}_{a},{\beta}_{e},b_{1},b_{2})\,E_{f}(t_{f})\,
  \Big\{ {\phi}_{\Upsilon}^{v}(x_{1}) \Big[
  {\phi}_{B_{c}}^{p}(x_{2})\, {\phi}_{D_{q}}^{p}(x_{3})\,
  r_{2}\,r_{3}\,(\bar{x}_{3}-x_{2})
   \nonumber \\ & &
 +{\phi}_{B_{c}}^{a}(x_{2})\, {\phi}_{D_{q}}^{a}(x_{3})\,
   \{ \frac{s\,(\bar{x}_{1}-x_{2})}{m_{1}^{2}}
   +2\,r_{3}^{2}\,(x_{3}-x_{1}) \} \Big]
   \nonumber \\ & &
   -r_{b}\,{\phi}_{\Upsilon}^{t}(x_{1})\,
   {\phi}_{B_{c}}^{a}(x_{2})\, {\phi}_{D_{q}}^{a}(x_{3}) \Big\}
   \label{amp-figf-01}, \\
  {\cal A}_{f}^{LR} &=& \frac{1}{N}
  {\int}_{0}^{1}dx_{1} {\int}_{0}^{1}dx_{2} {\int}_{0}^{1}dx_{3}
  {\int}_{0}^{\infty}b_{1} db_{1} {\int}_{0}^{\infty}b_{2} db_{2}
  {\int}_{0}^{\infty} db_{3}\, {\delta}(b_{2}-b_{3})\,{\alpha}_{s}(t_{f})
   \nonumber \\ & &
  H_{ef}({\alpha}_{a},{\beta}_{e},b_{1},b_{2})\,E_{f}(t_{f})\,
   \Big\{ {\phi}_{\Upsilon}^{v}(x_{1}) \Big[
  {\phi}_{B_{c}}^{p}(x_{2})\, {\phi}_{D_{q}}^{p}(x_{3})\,
   r_{2}\,r_{3}\, (\bar{x}_{3}-x_{2})
   \nonumber \\ & &
 +{\phi}_{B_{c}}^{a}(x_{2})\, {\phi}_{D_{q}}^{a}(x_{3})\,
   \{ \frac{s\,(x_{1}-x_{3})}{m_{1}^{2}}
   +2\,r_{2}^{2}\,(x_{2}-\bar{x}_{1}) \} \Big]
   \nonumber \\ & &
   +r_{b}\,{\phi}_{\Upsilon}^{t}(x_{1})\,
   {\phi}_{B_{c}}^{a}(x_{2})\, {\phi}_{D_{q}}^{a}(x_{3}) \Big\}
   \label{amp-figf-02}, \\
  {\cal A}_{f}^{SP} &=& \frac{1}{N}
  {\int}_{0}^{1}dx_{1} {\int}_{0}^{1}dx_{2} {\int}_{0}^{1}dx_{3}
  {\int}_{0}^{\infty}b_{1} db_{1} {\int}_{0}^{\infty}b_{2} db_{2}
  {\int}_{0}^{\infty} db_{3}\, {\delta}(b_{2}-b_{3})\,{\alpha}_{s}(t_{f})
   \nonumber \\ & &
   H_{ef}({\alpha}_{a},{\beta}_{e},b_{1},b_{2})\, E_{f}(t_{f})\,
   \Big\{ {\phi}_{\Upsilon}^{t}(x_{1})\, \Big[
  {\phi}_{B_{c}}^{a}(x_{2})\, {\phi}_{D_{q}}^{p}(x_{3})\,
  r_{3}\,(x_{1}-x_{3})
   \nonumber \\ & &
  + {\phi}_{B_{c}}^{p}(x_{2})\, {\phi}_{D_{q}}^{a}(x_{3})\,
  r_{2}\,(x_{2}-\bar{x}_{1}) \Big] +
  {\phi}_{\Upsilon}^{v}(x_{1})\, r_{b}\,
   \Big[ {\phi}_{B_{c}}^{a}(x_{2})\, {\phi}_{D_{q}}^{p}(x_{3})\,r_{3}
   \nonumber \\ & &
 +{\phi}_{B_{c}}^{p}(x_{2})\, {\phi}_{D_{q}}^{a}(x_{3})\,r_{2} \Big] \Big\}
   \label{amp-figf-03},
   \end{eqnarray}
   \begin{eqnarray}
  {\cal A}_{g}^{LL} &=& {\cal A}_{g}^{LR} =
  {\int}_{0}^{1}dx_{2} {\int}_{0}^{1}dx_{3}
  {\int}_{0}^{\infty}b_{2} db_{2}
  {\int}_{0}^{\infty}b_{3} db_{3}\,
  H_{gh}({\alpha}_{a},{\beta}_{g},b_{2},b_{3})\,
  E_{f}(t_{g})
   \nonumber \\ & &
  {\alpha}_{s}(t_{g})\,
   \Big\{ {\phi}_{B_{c}}^{a}(x_{2})\, {\phi}_{D_{q}}^{a}(x_{3})\,
   (x_{2}+r_{3}^{2}\,\bar{x}_{2})-2\,
   {\phi}_{B_{c}}^{p}(x_{2})\, {\phi}_{D_{q}}^{p}(x_{3})\,
   r_{2}\,r_{3}\,\bar{x}_{2} \Big\}
   \label{amp-figg}, \\
  {\cal A}_{h}^{LL} &=& {\cal A}_{h}^{LR} =
  {\int}_{0}^{1}dx_{2} {\int}_{0}^{1}dx_{3}
  {\int}_{0}^{\infty}b_{2} db_{2}
  {\int}_{0}^{\infty}b_{3} db_{3}\,
  H_{gh}({\alpha}_{a},{\beta}_{h},b_{3},b_{2})\,
  E_{h}(t_{h})
   \nonumber \\ & &
  {\alpha}_{s}(t_{h})\,
   \Big\{ {\phi}_{B_{c}}^{a}(x_{2})\, {\phi}_{D_{q}}^{a}(x_{3})\,
  (\bar{x}_{3}+r_{2}^{2}\,x_{3})
  +{\phi}_{B_{c}}^{a}(x_{2})\, {\phi}_{D_{q}}^{p}(x_{3})\,
  r_{3}\,r_{b}
  \nonumber \\ & &
 -2\,{\phi}_{B_{c}}^{p}(x_{2})\, {\phi}_{D_{q}}^{a}(x_{3})\,
  r_{2}\,r_{b}
 -2\,{\phi}_{B_{c}}^{p}(x_{2})\, {\phi}_{D_{q}}^{p}(x_{3})\,
  r_{2}\,r_{3}\,x_{3}  \Big\}
   \label{amp-figh},
   \end{eqnarray}
  where the mass ratio $r_{i}$ $=$ $m_{i}/m_{1}$;
  $\bar{x}_{i}$ $=$ $1$ $-$ $x_{i}$;
  variable $x_{i}$ is the longitudinal momentum fraction
  of the valence quark;
  $b_{i}$ is the conjugate variable of the
  transverse momentum $k_{i{\perp}}$;
  and ${\alpha}_{s}(t)$ is the QCD coupling at the
  scale of $t$.

  The function $H_{i}$ are defined as follows.
   \begin{eqnarray}
   H_{ab}({\alpha}_{e},{\beta},b_{i},b_{j})
   &=& K_{0}(\sqrt{-{\alpha}_{e}}b_{i})
   \Big\{ {\theta}(b_{i}-b_{j})
   K_{0}(\sqrt{-{\beta}}b_{i})
   I_{0}(\sqrt{-{\beta}}b_{j})
   + (b_{i}{\leftrightarrow}b_{j}) \Big\}
   \label{hab}, \\
   H_{cd}({\alpha}_{e},{\beta},b_{2},b_{3}) &=&
   \Big\{ {\theta}(-{\beta}) K_{0}(\sqrt{-{\beta}}b_{3})
  +\frac{{\pi}}{2} {\theta}({\beta}) \Big[
   iJ_{0}(\sqrt{{\beta}}b_{3})
   -Y_{0}(\sqrt{{\beta}}b_{3}) \Big] \Big\}
   \nonumber \\ &{\times}&
   \Big\{ {\theta}(b_{2}-b_{3})
   K_{0}(\sqrt{-{\alpha}_{e}}b_{2})
   I_{0}(\sqrt{-{\alpha}_{e}}b_{3})
   + (b_{2}{\leftrightarrow}b_{3}) \Big\}
   \label{hcd}, \\
   H_{ef}({\alpha}_{a},{\beta},b_{1},b_{2}) &=&
   \Big\{ {\theta}(-{\beta}) K_{0}(\sqrt{-{\beta}}b_{1})
  +\frac{{\pi}}{2} {\theta}({\beta}) \Big[
   iJ_{0}(\sqrt{{\beta}}b_{1})
   -Y_{0}(\sqrt{{\beta}}b_{1}) \Big] \Big\}
   \nonumber \\ & & \!\!\!\!\!\!\!\!\!\!\!\!\!\!\!\!\!\!\!\!\!\!\!\!
  {\times} \frac{{\pi}}{2} \Big\{ {\theta}(b_{1}-b_{2})
   \Big[ iJ_{0}(\sqrt{{\alpha}_{a}}b_{1})
   -Y_{0}(\sqrt{{\alpha}_{a}}b_{1}) \Big]
   J_{0}(\sqrt{{\alpha}_{a}}b_{2})
   + (b_{1}{\leftrightarrow}b_{2}) \Big\}
   \label{hef}, \\
  H_{hg}({\alpha}_{a},{\beta},b_{i},b_{j}) &=&
  \frac{{\pi}^{2}}{4}
  \Big\{ iJ_{0}(\sqrt{{\alpha}_{a}}b_{j})
   -Y_{0}(\sqrt{{\alpha}_{a}}b_{j}) \Big\}
   \nonumber \\ &{\times}&
   \Big\{ {\theta}(b_{i}-b_{j})
   \Big[ iJ_{0}(\sqrt{{\beta}}b_{i})
   -Y_{0}(\sqrt{{\beta}}b_{i}) \Big]
   J_{0}(\sqrt{{\beta}}b_{j})
   + (b_{i}{\leftrightarrow}b_{j}) \Big\}
   \label{hgh},
   \end{eqnarray}
  where $J_{0}$ and $Y_{0}$ ($I_{0}$ and $K_{0}$) are the
  (modified) Bessel functions of the first and second kind,
  respectively;
  ${\alpha}_{e}$ (${\alpha}_{a}$) is the
  gluon virtuality of the emission (annihilation)
  diagrams;
  the subscript of the quark virtuality ${\beta}_{i}$
  corresponds to the indices of Fig.\ref{fig:fey}.
  The definition of the particle virtuality is
  listed as follows.
   \begin{eqnarray}
  {\alpha}_{e} &=& \bar{x}_{1}^{2}m_{1}^{2}
                +  \bar{x}_{2}^{2}m_{2}^{2}
                -  \bar{x}_{1}\bar{x}_{2}t
   \label{gluon-q2-e}, \\
  {\alpha}_{a} &=& x_{2}^{2}m_{2}^{2}
                +  \bar{x}_{3}^{2}m_{3}^{2}
                +  x_{2}\bar{x}_{3}s
   \label{gluon-q2-a}, \\
  {\beta}_{a} &=& m_{1}^{2} - m_{b}^{2}
               +  \bar{x}_{2}^{2}m_{2}^{2}
               -  \bar{x}_{2}t
   \label{beta-fa}, \\
  {\beta}_{b} &=& m_{2}^{2} - m_{c}^{2}
               +  \bar{x}_{1}^{2}m_{1}^{2}
               -  \bar{x}_{1}t
   \label{beta-fb}, \\
  {\beta}_{c} &=& x_{1}^{2}m_{1}^{2}
               +  x_{2}^{2}m_{2}^{2}
               +  \bar{x}_{3}^{2}m_{3}^{2}
   \nonumber \\ &-&
                  x_{1}x_{2}t
               -  x_{1}\bar{x}_{3}u
               +  x_{2}\bar{x}_{3}s
   \label{beta-fc}, \\
  {\beta}_{d} &=& x_{1}^{2}m_{1}^{2}
               +  x_{2}^{2}m_{2}^{2}
               +  x_{3}^{2}m_{3}^{2}
               -  m_{c}^{2}
    \nonumber \\ &-&
                  x_{1}x_{2}t
               -  x_{1}x_{3}u
               +  x_{2}x_{3}s
   \label{beta-fd}, \\
  {\beta}_{e} &=& x_{1}^{2}m_{1}^{2}
               +  x_{2}^{2}m_{2}^{2}
               + \bar{x}_{3}^{2}m_{3}^{2}
               -  m_{b}^{2}
   \nonumber \\ &-&
                  x_{1}x_{2}t
               -  x_{1}\bar{x}_{3}u
               +  x_{2}\bar{x}_{3}s
   \label{beta-fe}, \\
  {\beta}_{f} &=& \bar{x}_{1}^{2}m_{1}^{2}
               +  x_{2}^{2}m_{2}^{2}
               +  \bar{x}_{3}^{2}m_{3}^{2}
               -  m_{b}^{2}
   \nonumber \\ &-&
                  \bar{x}_{1}x_{2}t
               -  \bar{x}_{1}\bar{x}_{3}u
               +  x_{2}\bar{x}_{3}s
   \label{beta-ff}, \\
  {\beta}_{g} &=& x_{2}^{2}m_{2}^{2}
               +  m_{3}^{2}
               +  x_{2}s
   \label{beta-fg}, \\
  {\beta}_{h} &=& \bar{x}_{3}^{2}m_{3}^{2}
               +  m_{2}^{2}
               +  \bar{x}_{3}s
               - m_{b}^{2}
   \label{beta-fh}.
   \end{eqnarray}

  The typical scale $t_{i}$ and the Sudakov factor $E_{i}$
  are defined as follows, where the subscript $i$ corresponds
  to the indices of Fig.\ref{fig:fey}.
   \begin{eqnarray}
   t_{a(b)} &=& {\max}(\sqrt{-{\alpha}_{e}},\sqrt{-{\beta}_{a(b)}},1/b_{1},1/b_{2})
   \label{tab}, \\
   t_{c(d)} &=& {\max}(\sqrt{-{\alpha}_{e}},\sqrt{{\vert}{\beta}_{c(d)}{\vert}},1/b_{2},1/b_{3})
   \label{tcd}, \\
   t_{e(f)} &=& {\max}(\sqrt{{\alpha}_{a}},\sqrt{{\vert}{\beta}_{e(f)}{\vert}},1/b_{1},1/b_{2})
   \label{tef}, \\
   t_{g(h)} &=& {\max}(\sqrt{{\alpha}_{a}},\sqrt{{\beta}_{g(h)}},1/b_{2},1/b_{3})
   \label{tgh},
   \end{eqnarray}
   \begin{equation}
   E_{i}(t) =
   \Bigg\{ \begin{array}{lll}
  {\exp}\{ -S_{B_{c}}(t) \}, &~& i=a,b \\
  {\exp}\{ -S_{B_{c}}(t)-S_{D_{q}}(t) \}, & & i=c,d,e,f,g,h
   \end{array}
   \label{sudakov-exp}
   \end{equation}
   \begin{eqnarray}
   S_{B_{c}}(t) &=&
   s(x_{2},p_{2}^{+},1/b_{2})
  +2{\int}_{1/b_{2}}^{t}\frac{d{\mu}}{\mu}{\gamma}_{q}
   \label{sudakov-bc}, \\
   S_{D_{q}}(t) &=&
   s(x_{3},p_{3}^{+},1/b_{3})
  +2{\int}_{1/b_{3}}^{t}\frac{d{\mu}}{\mu}{\gamma}_{q}
   \label{sudakov-ds},
   \end{eqnarray}
  where ${\gamma}_{q}$ $=$ $-{\alpha}_{s}/{\pi}$ is the
  quark anomalous dimension;
  the explicit expression of $s(x,Q,1/b)$ can be found in
  the appendix of Ref.\cite{pqcd1}.
  \end{appendix}

  
  \end{document}